
\magnification=1200
\font\open=msbm10 
\def\mbox#1{{\leavevmode\hbox{#1}}}
\def\hspace#1{{\phantom{\mbox#1}}}
\def\oR{\mbox{\open\char82}}

\def\rS{{\rm S}}

\def\al{\alpha}
\def\be{\beta}

\def\Ga{\Gamma}

\def\ep{\epsilon}

\def\la{\lambda}

\def\om{\omega}
\def\Om{\Omega}

\def\Si{\Sigma}

\def\ze{\zeta}
\def\De{\Delta}

\def\Det{{\rm Det\,}}
\def\Real{{\rm Re\,}}

\def\zf{$\zeta$--function}
\def\zfs{$\zeta$--functions}


\def\frac#1/#2{\leavevmode\kern.1em
\raise.5ex\hbox{\the\scriptfont0 #1}\kern-.1em/\kern-.15em
\lower.25ex\hbox{\the\scriptfont0 #2}}
\def\sfrac#1/#2{\leavevmode\kern.1em
\raise.5ex\hbox{\the\scriptscriptfont0 #1}\kern-.1em/\kern-.15em
\lower.25ex\hbox{\the\scriptscriptfont0 #2}}

\def\noin{\noindent}

\def\comb#1#2{{\left(#1\atop#2\right)}}

\def\etc{{\it etc. }}

\def\eg{{\it e.g. }}
\def\ie{{\it i.e. }}
\def\cf{{\it cf }}
\def\pa{\partial}

\def\sumstar#1{{\mathop{{\sum}^*_{#1}}}}

\def\man{{\cal M}}
\def\can{{\cal N}}


\def\aop#1#2#3{{\it Ann. Phys.} {\bf {#1}} (19{#2}) #3}

\def\cmp#1#2#3{{\it Comm. Math. Phys.} {\bf {#1}} (19{#2}) #3}
\def\cqg#1#2#3{{\it Class. Quant. Grav.} {\bf {#1}} (19{#2}) #3}

\def\jmp#1#2#3{{\it J. Math. Phys.} {\bf {#1}} (19{#2}) #3}
\def\jpa#1#2#3{{\it J. Phys.} {\bf A{#1}} (19{#2}) #3}

\def\np#1#2#3{{\it Nucl. Phys.} {\bf B{#1}} (19{#2}) #3}
\def\pl#1#2#3{{\it Phys. Lett.} {\bf {#1}} (19{#2}) #3}
\def\pm#1#2#3{{\it Phil.Mag.} {\bf {#1}} ({#2}) #3}

\def\pr#1#2#3{{\it Phys. Rev.} {\bf {#1}} (19{#2}) #3}
\def\prA#1#2#3{{\it Phys. Rev.} {\bf A{#1}} (19{#2}) #3}

\def\prD#1#2#3{{\it Phys. Rev.} {\bf D{#1}} (19{#2}) #3}

\def\zfp#1#2#3{{\it Z. f. Phys.} {\bf {#1}} (19{#2}) #3}

\def\prs#1#2#3{{\it Proc. Roy. Soc.} {\bf A{#1}} (19{#2}) #3}
\def\pcps#1#2#3{{\it Proc. Camb. Phil. Soc.} {\bf{#1}} (19{#2}) #3}
\def\acm#1#2#3{{\it Acta Math.} {\bf {#1}} (19{#2}) #3}

\def\cpam#1#2#3{{\it Comm. pure and Appl. Math.} {\bf {#1}} (19{#2}) #3}

\def\dmj#1#2#3{{\it Duke Math. J.} {\bf {#1}} (19{#2}) #3}

\def\jdg#1#2#3{{\it J. Diff. Geom.} {\bf {#1}} (19{#2}) #3}
\def\jfa#1#2#3{{\it J. Func. Anal.} {\bf {#1}} (19{#2}) #3}

\def\ma#1#2#3{{\it Math. Ann.} {\bf {#1}} ({#2}) #3}
\def\mz#1#2#3{{\it Math. Zeit.} {\bf {#1}} ({#2}) #3}
\def\pams#1#2#3{{\it Proc. Am. Math. Soc.} {\bf {#1}} (19{#2}) #3}

\def\qjm#1#2#3{{\it Quart. J. Math.} {\bf {#1}} (19{#2}) #3}

\def\tams#1#2#3{{\it Trans. Am. Math. Soc.} {\bf {#1}} (19{#2}) #3}

\vskip 5truept
\rightline{MUTP/95/17}
\centerline{\bf FUNCTIONAL DETERMINANTS}
\centerline{\bf ON M\"OBIUS CORNERS}\footnote{}{Talk presented at the 3rd
Workshop on
Quantum Field Theory under the Influence of External Conditions, Leipzig,
September 1995.}
\vskip 15truept
\centerline{J.S.Dowker}
\vskip10truept
\centerline{\it Department of Theoretical Physics,}
\centerline{\it The University of Manchester,}
\centerline{\it Manchester, England.}
\vskip20truept
\noin{\bf 1. Historical remarks.}

My title and subject matter are entirely appropriate for a talk delivered in
Leipzig where M\"obius spent most of his distingushed academic career.
Furthermore, the fundamental quantum field theory paper by Heisenberg and
Euler originated here and the classic book by Pockels on the Helmholtz
equation was published in 1891 by Teubner's in their famous Handbuch series.
\vskip10truept
\noin {\bf 2. Brief motivation.}

A functional determinant is the determinant of an operator, thus extending
the standard notion from the finite to the infinite dimensional situation.
The reason for its physical importance is that the one-loop effective action,
or the effective action in a background field, is determined, in the
simplest case, by the determinant of the propagating (differential)
operator.

In such, or equivalent,
language this result dates to Feynman, Neumann and Schwinger in the
40's and 50's although the idea, and evaluation, of an effective action
is much older and can be traced back to the beginnings of quantum field
theory (QED) being associated with Uehling, Serber, Heisenberg \& Euler and
Weisskopf. Nowadays, the most popular way of doing things in field,
instanton and string theory is via path-integration.

If the background field is a gravitational one, the determinant is a
function of the metric and therein lies its mathematical importance.
Usually the operator is a geometric one such as the Laplacian, and the
determinant is a {\it geometric invariant}. In fact the alternating sum
(with coefficients) of form determinants on a manifold  is a
{\it topological} invariant -- the {\it analytic torsion}.

Functional determinants have been employed in connection with the
uniformisation theorem of Riemann surfaces and with the isospectral
problem. The functional determinant depends on, and is determined by, the
eigenvalues of the operator. It is thus a {\it spectral invariant}.
\vskip10truept
\noin{\bf 3. The zeta function.}

A convenient but by no means the only way of organising the eigenvalues is
via the Minakshisundaram \zf,
$$
\ze(s)=\sum{1\over\la^s}
$$
so that formally ({\it \`a la} Euler),
$$
\Det D=\prod\la=\exp\big({\scriptstyle\sum}\ln\la\big)=e^{-\ze'(0)}
$$
and I will hereafter consider the problem of finding the functional
determinant as synonymous with finding $\ze'(0)$. My emphasis is already
turning to the mathematical side.

The central quantity in this approach is thus the \zf. The sum definition
converges only for $s$ is a certain region, typically $s>d/2$ where $d$
is the dimension of the manifold $\man$, if $D$ is the Laplacian for example.
This makes a continuation to $s=0$ (and may be to other values) necessary.

One more standard thing. The heat-kernel, $K$ (the propertime kernel of
Fock, Nambu, Feynman and Schwinger), and the \zf\ are related by a Mellin
transform and contain the same information, modulo zero modes. Some formulae
will follow.

The coefficients in the short-time expansion of $K$ are important quantities
determining the divergences of the field theory and its scaling behaviour. If
$D$ is a geometric operator, they are geometric invariants. Some explicit
general forms exist for the early coefficients and specific algorithms for
determining all the coefficients can be found in the contribution by
Schimming to this conference.

The analytic structure, \ie the poles, and the values of the \zf\ at specific
$s$, is related to these coefficients whose values in any particular
case can then be found from the \zf. They can be checked against the
general forms, or, more usefully, can be used in the  determination of
any unknown coefficients.

The calculations I shall later report on are to be
taken in this spirit of `Special Case Evaluation'.

We will hear more things about the \zf\ in the following talk by Elizalde.

\vskip 10truept
\noin {\bf 4. Summary of methods.}

A number of possible approaches are available for finding the \zf.

First, if the eigenvalues are known explicitly, \eg $\la=n$, one might
look at $\sum(1/\la^s)$ and seek a continuation, often by
reducing the sum, in a more or less direct way, to known (\ie named) \zfs,
(in the above case to the Riemann \zf, $\ze_R(s)$) for which the
continuation has already been done for us. This would be the ideal
situation, if all we want is the answer. Alternatively, there might be
special function properties,
contour representations and summation formulae that can be used,
requiring a certain amount of ingenuity to effect the continuation.

Second, if the eigenvalues are known only {\it implicitly}, the above
approach is not possible and one might not be able to find $\ze(s)$ for
all $s$. However, sufficient information might still be available for our
purposes. This is the situation I want to consider in a specific case later.

I will introduce such a method via the modified \zf\
$$
\ze(s,m^2)=\sum{1\over(\la+m^2)^s}
$$
which bears the same relation to $\ze(s)$ that the Hurwitz \zf\ does to
the Riemann one and
$$\eqalign{
\ze(s,m^2)&={1\over\Ga(s)}\int_0^\infty\tau^{s-1}
e^{-m^2\tau}K(\tau)\,d\tau\cr
\noalign{\vskip10truept}
&=\int_{c-i\infty}^{c+i\infty}ds'(m^2)^{s-s'}{\Ga(s')\Ga(s-s')\over\Ga(s)}
\ze(s',0),\cr}
$$
$d/2<c<(d+1)/2$. (Dikii.)

$\ze(s,m^2)$ can be evaluated at an $s$, say $q$, where the sum
converges, the variable $m^2$ then providing the access to the required
information. (In this sense the construction of $\ze(q,m^2)$ is an
alternative regularisation process to that of continuing $\ze(s)$. Dikii,
Gelfand).

In physical terms, $m^2$ is a mass and, as such, was separated by
Feynman in his proper-time approach. In the heat-kernel, the mass produces
convergence at the upper, infinite limit of the Mellin transform allowing
one to substitute the short-time expansion of $K$. This yields an
alternative method of deducing the analytic properties of the \zf\ and
results in an asymptotic expansion of $\ze(s,m^2)$, valid for large $m^2$,
whose coefficients are obviously the heat-kernel expansion coefficients,
and provides
a method of finding these coefficients (Dikii, Moss). I am not concerned
with this aspect here, being more interested in the functional
determinant and will present a general method of obtaining this that also
involves the asymptotic behaviour as $m\to\infty$.
\vskip10truept
\noin{\bf5. Weierstrass regularisation.}

The first step is to regularise the sum definition of $\ze(s,m^2)$ in the
following way
$$
\ze^*(s,m^2)=\sumstar{}{1\over(\la+m^2)^s}\equiv
\sum\bigg({1\over(\la+m^2)^s}-{1\over\la^s}-\sum_{k=1}^M\comb{-s}k
{m^{2k}\over\la^{k+s}}\bigg)
$$
where sufficient terms in the Taylor series, for any particular value
of $s$, have been removed to ensure convergence. I refer to this as
{\it Weierstrass regularisation}, the reason being that if the expression
is differentiated with respect to $s$, and $s$ set to zero I find
$$
\ze^{*'}(0,m^2)=-\sum\bigg(\ln\big(1+m^2/\la\big)+P(m^2/\la)\bigg)
=-\sumstar{}\ln\big(1+m^2/\la\big)
$$
where the polynomial $P$ is
$$
P(x)=x+{x^2\over2}+\ldots+{x^{2[d/2]}\over{[d/2]}}
$$
and we recognise
$$
e^{-\ze^{*'}(0,m^2)}=\prod\bigg(1+{m^2\over\la}\bigg)e^{P(m^2/\la)}
$$
as a {\it Weierstrass canonical product}. In quantum field theory, a
modified determinant occurs already in the work of Schwinger.

Let me now perform the summation over $\la$ in $\ze^{*}(s,m^2)$
to give the continuation
$$
\ze^*(s,m^2)=\ze(s,m^2)-\ze(s,0)-\sum_{k=1}^M\comb{-s}k
m^{2k}\ze(s+k,0)\bigg).
$$
In particular
$$
\ze^{*'}(0,m^2)=\ze'(0,m^2)-\ze'(0,0)
-{\pa\over\pa s}\sum_{k=1}^{[d/2]}\comb{-s}km^{2k}\ze(s+k,0)\bigg|_{s=0}.
$$

Combining these two expressions for $\ze^{*'}(0,m^2)$, and turning the
equation around gives the quantity I am seeking,
$$
\ze'(0,0)=\ze'(0,m^2)+\sumstar{}\ln\big(1+m^2/\la\big)-
{\pa\over\pa s}\sum_{k=1}^{[d/2]}\comb{-s}km^{2k}\ze(s+k,0)\bigg|_{s=0}.
$$
(\cf Voros, Quine, Heydari \& Song, Jorgenson \& Lang.) The last term can be
carried further, but I don't need to here.

The essential, rather trivial, point now is that the right-hand side of
this equation
must be mass-independent and so can be calculated at any convenient value
of $m$, in particular at $m=\infty$. It is not necessary to check
that all $m$-dependence does disappear. One just needs to keep the constant,
mass-independent parts of the various asymptotic limits. This is the method.

The asymptotic expansion of $\ze'(0,m^2)$ referred to earlier shows that
$$
\ze'(0,m^2)\sim 0+O(\ln m^2)
$$
then I arrive at the final, working formula
$$
\ze'(0)\approx\lim_{m\to\infty}\sumstar{}\ln\big(1+m^2/\la\big)
$$or
$$
\lim_{m\to\infty}\sumstar{}\ln\big(1+m^2/\la\big)=\ze'(0)+O(\ln m).
$$

This is a perfectly general equation. I want to illustrate the method by
applying it to the situation where the eigenvalues are known only
implicitly. Instead of giving a general treatment, I proceed now to the
specific calculation I wish to report on at this meeting.

I should say that related but different, methods have been developed
by Barvinsky, Kamenshchik, Karmazin and Mishakov and by Bordag, Geyer,
Kirsten and Elizalde, a group based mostly here at Leipzig.
\vskip 10truept
\noindent{\bf 6. The geometry and the calculation.}

Apart from any physical motivation, the plan is to look for geometrical
domains on which there is sufficient knowledge of the eigenvalue problem
to enable the functional determinant to be evaluated in `closed form'.
The operator is the Laplacian on scalars. Other fields can be treated
of course but I just want to illustrate the technique.
Pockels' book contains valuable information on this eigenvalue problem.
(Incidentally, this book contains probably the first discussion of the
eigenvalue counting function.)

The basic geometry I have in mind is that of the
{\it Euclidean $(d+1)$--ball}.Although I have been at
pains to say that I have no immediate physical motivation in mind, I should
point out that a considerable amount of work has been performed on such, and
related, geometries in the context of quantum cosmology by \eg D'Eath,
Esposito, Schleich, Moss, Pollifrone, Vassilevich and the Russian group, so
my calculation may not be entirely sterile.

The functional determinants (and heat-kernel coefficients) on the full ball
have been determined by the Leipzig group in the scalar case, and by
Kirsten and Cognola for other fields using methods that parallel, but
are in detail different to, my own. Of course, I like my technique better.

Actually I will not be dealing with the full ball here, but rather with a
{\it bounded, generalised cone} of a special sort.
This is, loosely speaking, the solid angle
subtended at the origin of the ball by a finite domain on its surface.
It is not possible to treat arbitrary shapes, I think, and I shall be
restricted to domains with special symmetry associations which I will
describe in a moment.

The reason for the calculation is purely one of aesthetics and not governed
by any physical motivation. At the formal level, I wish to advertise the
usefulness of the Barnes $\ze$- and multiple $\Ga$-functions.

The bounded, generalized cone is defined as the space $\oR^+\times\can$ with
the hyperspherical metric
$$
ds^2=dr^2+r^2d\Si^2
$$where $d\Si^2$ is the metric on the manifold $\can$, and $r$ runs from $0$
to $1$. For us, $\can\sim \rS^d$ locally and $d\Si^2$ is the metric on the
$d$-sphere. (\cf Cheeger.)

To set up the formalism, I outline some very standard textbook material.
The Laplacian can be written
$$
\De={\pa^2\over\pa r^2}+{d\over r}{\pa\over\pa r}-{1\over r^2}\De_S
$$where $\De_S$ here is the Laplacian on the sphere.

The eigenmodes of $\De$, with eigenvalue $-\al^2$, are of the form
$$
{J_{\nu}(\al r)\over r^{(d-1)/2}}\,Y(\Om)
$$where the harmonics on $\can$ satisfy
$$
\De_S Y(\Om)=-\bar\la^2 Y(\Om)
$$
and
$$
\nu^2= \bar\la^2+(d-1)^2/4.
$$

At this point I restrict the region $\can$ to be a fundamental domain of the
complete symmetry group, $\Ga$, of a $d+1$-dimensional polytope acting on its
circumscribing sphere $\rS^d$. I have used such regions several times
before and, following Terras, have referred to them as {\it
M\"obius corners}, ${\cal C}({\bf d})$. In the present case, when the range of
$r$ is limited, the corner will be a bounded one. An interesting aspect of
these spaces is that their boundaries are only piecewise smooth.

In three dimensions, the infinite corner can be realised as three planar
mirrors intersecting at one point to form a {\it trihedral kaleidoscope}
(M\"obius' term). To qualify as a genuine M\"obius corner, the dihedral
angles must be such as to allow an integral number of similar corners to
fill out the entire $4\pi$ solid angle when arranged around a common point
of intersection.

Incidentally, if the domains are identified to form an orbifold, their edges
(running through the origin) form three angular defects (cosmic strings).
The Casimir effect has been calculated in the infinite corner.
(Dowker and Chang.)

The eigenfunctions and eigenvalues on $\can$ are  determined by group and
invariant theory. The usual spherical harmonics are filtered
(symmetry adapted) by the condition of being invariant under $\Ga$.
The important fact for my calculation is that the eigenvalues $\bar\la$ are
given by
$$
\bar\la^2_{\bf n}=({\bf n.d}+a)^2-(d-1)^2/4
$$
where ${\bf d}$ is a $d$-dimensional vector of the integer degrees
associated with the tiling group, $\Ga$, and where the integers $n_i$
range from zero to
infinity. The degeneracy of any particular eigenvalue depends on the number
of coincidences as ${\bf n}$ varies. The parameter $a$ determines the type
of boundary conditions on the {\it sides} of the cone and, in what is
effectively an image-based approach, I can, unfortunately, treat only
Dirichlet and Neumann conditions with $a$ equalling $\sum d_i -(d-1)/2$
and $(d-1)/2$ respectively.

You now see the nice circumstance that the order of the Bessel function
is still an integer or a half odd-integer, depending on the oddness or
evenness of the dimension $d$,
$$
\nu_{\bf n}={\bf n.d}+a.
$$

The boundary condition on the spherical end of the cone at $r=1$ has still
to be fixed. Just as for the full ball, one can impose Dirichlet, Neumann or
Robin conditions. For Dirichlet the implicit
eigenvalue condition is
$$
J_{{\bf n.d}+a}(\al)=0\,,
$$
while Robin boundary conditions yield,
$$
\be J_{{\bf n.d}+a}(\al)+\al J'_{{\bf n.d}+a}(\al)=0,
$$
where $\be$ is a constant. (Actually $\be$ could be a function on $\can$ but I
won't discuss this here.)

In order to extract the eigenvalue properties from these equations, I follow
previous workers and use Euler's representation of the Bessel function in
terms of its zeros, to write, for example,
$$
\sum_p\ln\big(2^p p!m^{-p}I_p(m)\big)
=\sum_{p,\al_p}\ln\bigg(1+{m^2\over\al_p^2}\bigg)
=\sum_\la\ln\bigg(1+{m^2\over\la}\bigg).
$$

Then
$$
\ze'(0)\approx\lim_{m\to\infty}\sumstar{p}\ln\big(2^p p!m^{-p}I_p(m)\big),
$$
where $p={\bf n.d}+a$ and the sum over $p$ means a multiple sum over
the integers ${\bf n}$.

The further analysis of this expression involves some algebra and will be of
interest only for those currently working in this field. However let me give
you a flavour of the calculation.

The asymptotic behaviour of the Bessel function is known (Olver) and the
$\ln p!$ term can be replaced by an integral representation. Then
$$\eqalign{
\ze'(0)&\sim\sumstar{\bf n} \bigg[p\ln{2p\over p+\ep}+(\ep-p)
-{1\over2}\ln{\ep\over p}+\sum_{l=1}^\infty{T_l(t)\over \ep^l}\cr
&\hspace{*********}+\int_0^\infty\bigg({1\over2}
-{1\over\tau}+{1\over e^\tau-1}\bigg){e^{-\tau p}\over\tau}\,d\tau\bigg],\cr}
\eqno(1)$$
where the $T_l(t)$ are Olver's polynomials in $t=p/\ep$ and
$\ep^2\equiv p^2+m^2$.

I will indicate how to deal with the first three terms. One simply applies
the Weierstrass regularisation again. Thus
$$\sumstar{{\bf n}}\,{p^N\over\ep^{2s}}=
\sum_{\bf n}p^N\bigg[{1\over\ep^{2s}}-{1\over p^{2s}}-
\sum_{h=1}^M\comb{-s}h{m^{2h}\over p^{2h+2s}}\bigg]
\eqno(2)$$ from which, if $s$ is given certain values, possibly after
differentiating, the required limits can be found.

I remind you that $p={\bf n.d}+a$. If the sums are done you can see that an
important role is played by the Barnes \zf\ defined by
$$\eqalign{
\zeta_d(s,a\mid{\bf {d}})&={i\Gamma(1-s)\over2\pi}\int_L d\tau {\exp(-a\tau)
(-\tau)^{s-1}\over\prod_{i=1}^d\big(1-\exp(-d_i\tau)\big)}\cr
&=\sum_{{\bf {n}}={\bf 0}}^\infty{1\over(a+{\bf {n.d}})^s},\qquad
\Real\, s>d.\cr}
$$

By general theory, the asymptotic limit of the first term in (2) is
$$
\sum_{\bf n}{1\over\big(({\bf n.d}+a)^2+m^2\big)^s}
\sim\sum_{k=0,1/2,1,\ldots}
(m^2)^{d/2-k-s}{\Ga(s-d/2+k)\over\Ga(s)}\,C_k
$$
where the $C_k$ are the coefficients in the short-time expansion of the
heat-kernel associated with the Barnes \zf,
$\ze_d\big(2s,a\mid{\bf d}\big)$ \eg
$$
(-1)^k k!C_{d/2+k}=\ze_d(-2k,a\mid{\bf d}),
$$
with
$$
\zeta_d(-q,a\mid{\bf d})={(-1)^dq!\over(d+q)!{\textstyle\prod}d_i}
B_{d+q}^{(d)}(a\mid{\bf d})
$$
in terms of generalized Bernoulli functions.

Actually I do not need these explicit expressions. Applying the Weierstrass
regularisation only the second term in (2) contributes and one easily obtains
the useful limits
$$\eqalign{
&\sum_{\bf n}p^N(\ep-p)\sim -\ze_d\big(-N-1,a\mid{\bf d}\big)+O(\ln m),\cr
&\sum_{\bf n}p^N\ln\big({2p\over p+\ep}\big)\sim
-\ze_d'\big(-N,a\mid{\bf d}\big)+\ln2\,
\ze_d\big(-N,a\mid{\bf d}\big)+O(\ln m),\cr
&\sum_{\bf n}p^N\ln\big({\ep\over p}\big)\sim
\ze_d'\big(-N,a\mid{\bf d}\big)+O(\ln m),\cr}
$$
enabling the first three terms in equation (1) to be found. I won't inflict
the remainder of the calculation on you, although a number of interesting
technical points arise. I will just write down the final
answer, which is reasonably compact,
$$\eqalign{
\ze_{\cal C(\bf d)}'(0)=\ze_{d+1}'&\big(0,a+1\mid{\bf d},1\big)+
\ln2\bigg(\ze_d\big(-1,a\mid{\bf d}\big)+
\sum_{l=1}^dT_l(1)\,N_l(d)\bigg)\cr
&+\sum_{l=1}^dN_l(d)\int_0^1t^{l-1}T_l''(t)\,dt
+{1\over2}\sum_{l=1}^dT_l(1)\,
N_l(d)\sum_{k=1}^{(l-1)/2}{1\over k}.\cr}
$$
The $N_l(d)$ are the residues of the Barnes \zf\ and, like the $T_l(1)$, are
given by Bernoulli functions.

Everything is quite explicit, apart from the derivative of the Barnes \zf\
which, in any particular case, could be reduced to a number of Hurwitz \zfs.
I have thus achieved my aim of finding a closed form. More explicit
expressions exist for the full sphere case and agree with those of the
Leipzig group when evaluated at specific dimensions.
\vskip 10truept
\noindent{\bf 7. The Robin case.}

A little more involved, and therefore more interesting, is the case of Robin
conditions on the spherical end of the cone. I am certainly not going to go
through the details systematically but I do want to point out the following
little bit of formalism. This is mostly self indulgence but some of you
might be interested.

A new term in this case is
$$
\sumstar{\bf n}\ln\bigg(1+{\be\over{{\bf n.d}+a}}\bigg)
$$
which is the Weierstrass product associated with the eigenvalues
${\bf n.d}+a$ and can be dealt with according to the general
result given earlier which, in this case, relates it to the derivative of the
Barnes \zf, $\ze_d'\big(0,a\mid{\bf d}\big)$. Applying this relation, I find
$$
\sumstar{\bf n}\ln\bigg(1+{\be\over{{\bf n.d}+a}}\bigg)=
\ln\bigg({\Ga_d(a)\over\Ga_d(a+\be)}\bigg)+\sum_{l=1}^d{\be^l\over l!}
\psi^{(l)}_d(a),
$$
in terms of the multiple $\Ga$- and $\psi$-functions. In fact this
generalises Barnes' product  for the multiple $\Ga$-function.

The power series
$$
\sumstar{\bf n}\ln\bigg(1+{\be\over{{\bf n.d}+a}}\bigg)=
\sum_{l=d}^\infty{(-1)^l\be^{l+1}\over{l+1}}\ze_d\big(l+1,a\mid{\bf d}\big)
$$
is also easily derived.

I give the final determinant expression for completeness,
$$\eqalign{
\ze_{\cal C(\bf d)}&'(0,\be)=\ze_{d+1}'\big(0,a\mid{\bf d},1\big)
-\ln\bigg({\Ga_d(a)\over\Ga_d(a+\be)}\bigg)-\sum_{l=2,\ldots}^d{\be^l
\over l}N_l(d)\sum_{k=1}^{l/2}{1\over2k-1}\cr
&+{1\over2}\sum_{l=1,3\ldots}^dR_l(\be,1)\,N_l(d)\sum_{k=1}^{(l-1)/2}
{1\over k}+\sum_{l=1}^dN_l(d)\int_0^1t^{l-1}R_l''(\be,t)\,dt\cr
&+\ln2\bigg(\ze_d\big(-1,a\mid
{\bf d}\big)+\sum_{l=1,3,\ldots}^dR_l(\be,1)\,N_l(d)\bigg).\cr}
$$
It is not much more complicated than the Dirichlet one. The $R(\be,t)$ are
the relevant Olver asymptotic polynomials (Moss).

Using the definition of the multiple $\Ga$-function and $\Ga$-modular form,
$\rho$, I can rewrite part of this expression,
$$\eqalign{
\ze_{d+1}'\big(0,a\mid{\bf d},1\big)
-\ln\bigg({\Ga_d(a)\over\Ga_d(a+\be)}\bigg)&=
\ln\bigg({\Ga_{d+1}(a)\over\rho_{d+1}({\bf d},1)}{\Ga_d(a+\be)
\over\Ga_d(a)}\bigg)\cr
&=\ln\bigg({\rho_d({\bf d})\over\rho_{d+1}({\bf d},1)}{\Ga_d(a+\be)
\over\Ga_{d+1}(a+1)}\bigg).\cr}
$$

Numerically, the difficulty lies in the evaluation of the multiple
$\Ga$-functions and the $\Ga$-modular forms. The definitions are,

$$\lim_{\ep\to0}\ze'_r\big(0,\ep\mid{\bf d}\big)
=-\ln\ep-\ln\rho_r({\bf d}),\quad\quad
\ze'_r\big(0,a\mid{\bf d}\big)=\ln\bigg({\Ga_r(a)\over\rho_r({\bf d})}
\bigg).
$$
\vskip 10truept
\noin{\bf8. Note on the Barnes \zf.}

The original papers of Barnes contain a lot of information about this
function, which can be looked upon as a multidimensional generalisation of
the Hurwitz \zf. It forms a happy playground for those of us amused by
special functions, recursion formulae, curious identities \etc

I will not attempt to summarize this material here. All I want to do is to
point out that the Barnes \zf\ arises `naturally' when considering the
$d$-dimensional harmonic oscillator with operator
$$
-\De_{\rm HO}=-\nabla^2+\sum_{i=1}^d\om_i^2 x_i^2
$$
whose eigenvalues are $\la=2{\bf n.\om}+\sum_i\om_i$. The corresponding
\zf\ is
$$
\ze_{\rm HO}(s)=\sum_{\bf n=0}^\infty{1\over\big(2{\bf n.\om}
+\sum\om_i\big)^s}=\ze_d\big(s,{\scriptstyle\sum}\om_i\mid 2\om\big).
$$
The diagonal heat-kernel can also be written down immediately
$$
K_{\rm HO}\big({\bf x},\tau;{\bf x},0\big)=\prod_i
{\exp\big(-\om_i x_i^2\tanh\om_i\tau\big)\over2\pi\sinh2\om_i\tau}.
$$

Since a constant magnetic field is more or less mathematically equivalent to
a multidimensional harmonic oscillator, the Barnes \zf\ will occur here also.
\vskip 10truept
\noin{\bf9. Conclusion.}

There is no conclusion because I am still pursuing the calculation, trying
to make sense of some special cases, and I haven't even mentioned Neumann
conditions or other fields like spin-1/2 and Maxwell.

The domains I have been engaged with form a discrete series and it would be
nice to have some continuously varying parameter so that a graph or two
could be drawn. It is possible to discuss a spherical wedge with an
arbitrary opening
angle (in which case one can have Robin conditions on the sides) and it may
be possible to treat a standard spherical ice-cream cone, \ie one whose
surface domain, $\can$, is a {\it spherical} $d$-ball or cap. The M\"obius
cone can also be truncated at an {\it inner} radius giving a portion of a
{\it shell}.
\vskip10truept
\noin{\bf10. Added note.}

I have learnt at this conference that Weierstrass regularisation has been
used by Wipf in an interesting discussion of tunnel determinants.
\vskip10truept
\noin{\bf References}
\vskip20truept

{E.W.Barnes {\it Trans. Camb. Phil. Soc.} {\bf 19} (1903)
374, 426.}

{A.O.Barvinsky, Yu.A.Kamenshchik and I.P.Karmazin \aop {219}
{92}{201}.}

{M.Bordag, E.Elizalde and K.Kirsten {\it Heat kernel
coefficients of the Laplace operator\break
\hspace{*****} on the D-dimensional ball}
 UB-ECM-PF 95/3; hep-th/9503023.}

{M.Bordag, B.Geyer, K.Kirsten and E.Elizalde {\it Zeta function
determinant of the \break
\hspace{*****}Laplace operator on the D-dimensional ball} UB-ECM-PF
95/10; hep-th /9505157.}

{J.Cheeger \jdg{18}{83}{575}.}

{J.Cheeger and M.Taylor \cpam{35}{82}{275,487}.}

{P.D.D'Eath and G.V.M.Esposito \prD{43}{91}{3234}.}

{P.D.D'Eath and G.V.M.Esposito \prD{44}{91}{1713}.}

{L.A.Dikii {\it Usp. Mat. Nauk.} {\bf13} (1958) 111.}

{J.S.Dowker {\it Robin conditions on the Euclidean ball}
MUTP/95/7; hep-th/9506042.}

{J.S.Dowker and Peter Chang \pr{D46}{92}{3458}.}

{J.S.Dowker {\it Spin on the 4-ball} MUTP/95/13; hep-th/9508082.}

W.Heisenberg and H.Euler \zfp{98}{36}{714}

J.Jorgenson and S.Lang Lect. Notes in Math. 1564 Springer-Verlag,
Berlin 1993.

{Yu.A.Kamenshchik and I.V.Mishakov \prD{47}{93}{1380}.}

{Yu.A.Kamenshchik and I.V.Mishakov {\it Int. J. Mod. Phys.}
{\bf A7} (1992) 3265.}

{K.Kirsten and G.Cognola {\it Heat-kernel coefficients and
functional determinants for
\hspace{*****}higher spin fields on the ball} UTF354. hep-th/9508088.}

{I.G.Moss \cqg{6}{89}{659}.}

{F.Pockels {\it \"Uber die partielle Differentialgleichung
$\Delta u+k^2u=0$}, B.G.Teubner, Leipzig
\hspace{*****}1891.}

{J.R.Quine, S.H.Heydari and R.Y.Song \tams{338}{93}{213}.}

J.Schwinger \pr {93}{53}{615}

{A.Voros \cmp{110}{87}{110}.}

A.Wipf \np{269}{86}{24}; and earlier references here.

\bye

\vskip 10truept
\noin{\bf{References}}
\vskip 5truept
\begin{putreferences}
\ref{Nielsen}{N.Nielsen {\it Handbuch der Theorie der Gamma Funktion},
Teubner, Leipzig, 1906.}
\ref{Milne-Thomson}{Milne-Thomson {\it Finite Differences}.}
\ref{Norlund}{N.E.N\"orlund \acm{43}{22}{121}.}
\ref{Rayleigh}{Lord Rayleigh{\it Theory of Sound} vols.I and II,
MacMillan, London, 1877,78.}
\ref{Donnelly} {H.Donnelly \ma{224}{1976}161.}
\ref{Fur2}{D.V.Fursaev {\sl Spectral geometry and one-loop divergences on
manifolds with conical singularities}, JINR preprint DSF-13/94,
hep-th/9405143.}
\ref{HandE}{S.W.Hawking and G.F.R.Ellis {\sl The large scale structure of
space-time} Cambridge University Press, 1973.}
\ref{FandM}{D.V.Fursaev and G.Miele \pr{D49}{94}{987}.}
\ref{BandH}{J.Br\"uning and E.Heintze \dmj{51}{84}{959}.}
\ref{Cheeger}{J.Cheeger \jdg{18}{83}{575}.}
\ref{CandT}{J.Cheeger and M.Taylor \cpam{35}{82}{275,487}.}
\ref{SandW}{K.Stewartson and R.T.Waechter \pcps{69}{71}{353}.}
\ref{CandJ}{H.S.Carslaw and J.C.Jaeger {\it The conduction of heat
in solids} Oxford, The Clarendon Press, 1959.}
\ref{BandH}{H.P.Baltes and E.M.Hilf {\it Spectra of finite systems}.}
\ref{Epstein}{P.Epstein \ma{56}{1903}{615}.}
\ref{Kennedy1}{G.Kennedy \pr{D23}{81}{2884}.}
\ref{Kennedy2}{G.Kennedy PhD thesis, Manchester (1978).}
\ref{Kennedy3}{G.Kennedy \jpa{11}{78}{L173}.}
\ref{Luscher}{M.L\"uscher, K.Symanzik and P.Weiss \np {173}{80}{365}.}
\ref{Polyakov}{A.M.Polyakov \pl {103}{81}{207}.}
\ref{Bukhb}{L.Bukhbinder, V.P.Gusynin and P.I.Fomin {\it Sov. J. Nucl.
 Phys.} {\bf 44} (1986) 534.}
\ref{Alvarez}{O.Alvarez \np {216}{83}{125}.}
\ref{KCD}{G.Kennedy, R.Critchley and J.S.Dowker \aop{125}{80}{346}.}
\ref{DandS}{J.S.Dowker and J.P.Schofield \jmp{31}{90}{808}.}
\ref{Dow1}{J.S.Dowker \cmp{162}{94}{633}.}
\ref{Dow2}{J.S.Dowker \cqg{11}{94}{557}.}
\ref{Dow3}{J.S.Dowker \jmp{35}{94}{4989}; erratum {\it ibid}, Feb.1995.}
\ref{Dow5}{J.S.Dowker {\it Heat-kernels and polytopes} To be published}
\ref{Dow6}{J.S.Dowker \pr{D50}{94}{6369}.}
\ref{Dow7}{J.S.Dowker \pr{D39}{89}{1235}.}
\ref{Dow8}{J.S.Dowker {\it Robin conditions on the Euclidean ball}
MUTP/95/7; hep-th\break/9506042.}
\ref{Dow9}{J.S.Dowker {\it Oddball determinants} MUTP/95/12; hep-th/9507096.}
\ref{Dow10}{J.S.Dowker \pr{D28}{83}{3013}.}
\ref{Dow11}{J.S.Dowker \jmp{30}{89}{770}.}
\ref{Dow12}{J.S.Dowker {\it Spin on the 4-ball} MUTP/95/13; hep-th/9508082.}
\ref{DandC1}{J.S.Dowker and R.Critchley \prD {13}{76}{3224}.}
\ref{DandC2}{J.S.Dowker and R.Critchley \prD {13}{76}{224}.}
\ref{DandK}{J.S.Dowker and G.Kennedy \jpa{11}{78}{895}.}
\ref{ChandD}{Peter Chang and J.S.Dowker \np{395}{93}{407}.}
\ref{DandC}{J.S.Dowker and Peter Chang \pr{D46}{92}{3458}.}
\ref{DandA}{J.S.Dowker and J.S.Apps \cqg{12}{95}{1363}.}
\ref{Dowkerccs}{J.S.Dowker \cqg{4}{87}{L157}.}
\ref{DandA2}{J.S.Dowker and J.S.Apps, {\it Functional determinants on certain
domains}. To appear in the Proceedings of the 6th Moscow Quantum Gravity
Seminar held in Moscow, June 1995; hep-th/9506204.}
\ref{BandG}{P.B.Gilkey and T.P.Branson \tams{344}{94}{479}.}
\ref{Schofield}{J.P.Schofield Ph.D.thesis, University of Manchester,
(1991).}
\ref{Barnesa}{E.W.Barnes {\it Trans. Camb. Phil. Soc.} {\bf 19} (1903)
374.}
\ref{Barnesb}{E.W.Barnes {\it Trans. Camb. Phil. Soc.} {\bf 19} (1903)
426.}
\ref{BandG2}{T.P.Branson and P.B.Gilkey {\it Comm. Partial Diff. Equations}
{\bf 15} (1990) 245.}
\ref{Pathria}{R.K.Pathria {\it Suppl.Nuovo Cim.} {\bf 4} (1966) 276.}
\ref{Baltes}{H.P.Baltes \prA{6}{72}{2252}.}
\ref{Spivak}{M.Spivak {\it Differential Geometry} vols III, IV, Publish
or Perish, Boston, 1975.}
\ref{Eisenhart}{L.P.Eisenhart {\it Differential Geometry}, Princeton
University Press, Princeton, 1926.}
\ref{Moss}{I.G.Moss \cqg{6}{89}{659}.}
\ref{Barv}{A.O.Barvinsky, Yu.A.Kamenshchik and I.P.Karmazin \aop {219}
{92}{201}.}
\ref{Kam}{Yu.A.Kamenshchik and I.V.Mishakov \prD{47}{93}{1380}.}
\ref{KandM}{Yu.A.Kamenshchik and I.V.Mishakov {\it Int. J. Mod. Phys.}
{\bf A7} (1992) 3265.}
\ref{DandE}{P.D.D'Eath and G.V.M.Esposito \prD{43}{91}{3234}.}
\ref{DandE2}{P.D.D'Eath and G.V.M.Esposito \prD{44}{91}{1713}.}
\ref{Rich}{K.Richardson \jfa{122}{94}{52}.}
\ref{Osgood}{B.Osgood, R.Phillips and P.Sarnak \jfa{80}{88}{148}.}
\ref{BCY}{T.P.Branson, S.-Y. A.Chang and P.C.Yang \cmp{149}{92}{241}.}
\ref{Vass}{D.V.Vassilevich.{\it Vector fields on a disk with mixed
boundary conditions} gr-qc /9404052.}
\ref{MandP}{I.Moss and S.Poletti \pl{B333}{94}{326}.}
\ref{Kam2}{G.Esposito, A.Y.Kamenshchik, I.V.Mishakov and G.Pollifrone
\prD{50}{94}{6329}.}
\ref{Aurell1}{E.Aurell and P.Salomonson \cmp{165}{94}{233}.}
\ref{Aurell2}{E.Aurell and P.Salomonson {\it Further results on functional
determinants of laplacians on simplicial complexes} hep-th/9405140.}
\ref{BandO}{T.P.Branson and B.\O rsted \pams{113}{91}{669}.}
\ref{Elizalde1}{E.Elizalde, \jmp{35}{94}{3308}.}
\ref{BandK}{M.Bordag and K.Kirsten {\it Heat-kernel coefficients of
the Laplace operator on the 3-dimensional ball} hep-th/9501064.}
\ref{Waechter}{R.T.Waechter \pcps{72}{72}{439}.}
\ref{GRV}{S.Guraswamy, S.G.Rajeev and P.Vitale {\it O(N) sigma-model as
a three dimensional conformal field theory}, Rochester preprint UR-1357.}
\ref{CandC}{A.Capelli and A.Costa \np {314}{89}{707}.}
\ref{IandZ}{C.Itzykson and J.-B.Zuber \np{275}{86}{580}.}
\ref{BandH}{M.V.Berry and C.J.Howls \prs {447}{94}{527}.}
\ref{DandW}{A.Dettki and A.Wipf \np{377}{92}{252}.}
\ref{Weisbergerb} {W.I.Weisberger \cmp{112}{87}{633}.}
\ref{Voros}{A.Voros \cmp{110}{87}{110}.}
\ref{Pockels}{F.Pockels {\it \"Uber die partielle Differentialgleichung
$\Delta u+k^2u=0$}, B.G.Teubner, Leipzig 1891.}
\ref{Kober}{H.Kober \mz{39}{1935}{609}.}
\ref{Watson2}{G.N.Watson \qjm{2}{31}{300}.}

\ref{Lamb}{H.Lamb \pm{15}{1884}{270}.}
\ref{EandR}{E.Elizalde and A.Romeo International J. of Math. and Phys.
{\bf13} (1994) 453}

\ref{Watson1}{G.N.Watson {\it Theory of Bessel Functions} Cambridge
University Press, Cambridge, 1944.}
\ref{BGKE}{M.Bordag, B.Geyer, K.Kirsten and E.Elizalde, {\it Zeta function
determinant of the Laplace operator on the D-dimensional ball} UB-ECM-PF
95/10; hep-th /9505157.}
\ref{MandO}{W.Magnus and F.Oberhettinger {\it Formeln und S\"atze}
Springer-Verlag, Berlin, 1948.}
\ref{Olver}{F.W.J.Olver {\it Phil.Trans.Roy.Soc} {\bf A247} (1954) 328.}
\ref{Hurt}{N.E.Hurt {\it Geometric Quantization in action} Reidel,
Dordrecht, 1983.}
\ref{Esposito}{G.Esposito {\it Quantum Gravity, Quantum Cosmology and
Lorentzian Geometry}, Lecture Notes in Physics, Monographs, Vol. m12,
Springer-Verlag, Berlin 1994.}
\ref{Louko}{J.Louko \prD{38}{88}{478}.}
\ref{Schleich} {K.Schleich \prD{32}{85}{1989}.}
\ref{BEK}{M.Bordag, E.Elizalde and K.Kirsten {\it Heat kernel
coefficients of the Laplace operator on the D-dimensional ball}
 UB-ECM-PF 95/3; hep-th/9503023.}
\ref{ELZ}{E.Elizalde, S.Leseduarte and S.Zerbini.}
\ref{BGV}{T.P.Branson, P.B.Gilkey and D.V.Vassilevich {\it The Asymptotics
of the Laplacian on a manifold with boundary} II, hep-th/9504029.}
\ref{Erdelyi}{A.Erdelyi,W.Magnus,F.Oberhettinger and F.G.Tricomi {\it
Higher Transcendental Functions} Vol.I McGraw-Hill, New York, 1953.}
\ref{Quine}{J.R.Quine, S.H.Heydari and R.Y.Song \tams{338}{93}{213}.}
\ref{Dikii}{L.A.Dikii {\it Usp. Mat. Nauk.} {\bf13} (1958) 111.}
\ref{DandH}{P.D.D'Eath and J.J.Halliwell \prd{35}{87}{1100}.}
\ref{KandC}{K.Kirsten and G.Cognola, {\it Heat-kernel coefficients and
functional determinants for higher spin fields on the ball} UTF354. Aug. 1995.}
\ref{Louko}{J.Louko \prD{38}{88}{478}.}
\ref{MandP}{I.G.Moss and S.J.Poletti \pl{B333}{94}{326}.}
\ref{MandP2}{I.G.Moss and S.J.Poletti \np{341}{90}{155}.}
\ref{Luck}{H.C.Luckock \jmp{32}{91}{1755}.}
\ref{Poletti}{S.J.Poletti \pl{B249}{90}{355}.}
\ref{gilk}{P.B.Gilkey {\it Invariant theory, the heat equation and the
Atiyah-Singer index theorem}, Publish or Perish, Wilmington, DE, 1984.}
\ref{deB}{Louis de Broglie {\it Probl\`emes de propagation guide\'es des
ondes electromagnetiques} 2me. \'Ed. Gauthier-Villars, Paris, 1951.}
\end{putreferences}
\bye